\documentstyle[12pt]{article}

\newcommand{\be}{\begin{equation}}
\newcommand{\ee}{\end{equation}}
\newcommand{\bea}{\begin{eqnarray}}
\newcommand{\eea}{\end{eqnarray}}

\def\G{{\cal G}}                         %
\def\M{{\cal M}}                         %
\def\A{{\cal A}}                         %
\def\R{{\bf R}}                          %
\begin{document}

\thispagestyle{empty}
\setcounter{page}{0}

\renewcommand{\thefootnote}{\fnsymbol{footnote}}
\fnsymbol{footnote}

\vspace{.4in}

\begin{center} \Large\bf 

The chiral WZNW phase space \\
as a quasi-Poisson space
\end{center}

\vspace{.1in}

\begin{center}
J.~Balog$^{(a)}$, L.~Feh\'er$^{(b)}$\protect\footnote{
Corresponding author's e-mail: lfeher@sol.cc.u-szeged.hu, 
phone/fax: (+36) 62 544 368.}
and L.~Palla$^{(c)}$ \\

\vspace{0.2in}

$^{(a)}${\em  Research Institute for Nuclear and Particle 
Physics,} \\
       {\em Hungarian Academy of Sciences,} \\
       {\em  H-1525 Budapest 114, P.O.B. 49, Hungary}\\

\vspace{0.2in}       

$^{(b)}${\em Department of Theoretical Physics,
        University of Szeged,} \\
       {\em  H-6726 Szeged, Tisza Lajos krt 84-86,
 Hungary }\\

\vspace{0.2in}   

$^{(c)}${\em  Institute for Theoretical Physics, 
Roland E\"otv\"os University,} \\
{\em  H-1117, Budapest, P\'azm\'any P. s\'et\'any 1 A-\'ep,  
Hungary }\\

\end{center}

\vspace{.2in}

\begin{center} \bf Abstract
\end{center}

{\parindent=25pt
\narrower\smallskip\noindent
It is explained that the chiral WZNW phase space is a 
quasi-Poisson space 
with respect to the `canonical' Lie quasi-bialgebra which 
is the classical 
limit of Drinfeld's quasi-Hopf deformation of the universal 
enveloping algebra.
This exemplifies the notion of quasi-Poisson-Lie symmetry 
introduced recently 
by Alekseev and Kosmann-Schwarzbach, 
and also permits us to generalize certain dynamical twists 
considered  previously in this example.
}

\vspace{11 mm}
\begin{center}
keywords: WZNW model, exchange algebra, Lie quasi-bialgebra, 
quasi-Poisson space

\end{center}
\vfill\eject

\section{Introduction}
\setcounter{equation}{0}

The WZNW model \cite{W} is perhaps the most important model 
of 2-dimensional 
conformal field theory due to its applications in string 
theory and to the
fact that many other interesting models can be understood 
as its reductions.
The reduction procedures, such as Hamiltonian reduction and 
the Goddard-Kent-Olive coset construction, rely on the affine 
Kac-Moody
symmetry of the model.  Besides this linear symmetry, the 
WZNW model also provides an arena in which non-linear,
 quantum group 
symmetries appear \cite{GRS}. 

One of the quantum group aspects of the model is that the
fusion rules of the WZNW primary fields  reflect \cite{FGP} 
certain truncated Clebsch-Gordan series for tensor 
products of 
representations of a Drinfeld-Jimbo quantum group $U_q(\G)$.
Another very interesting result \cite{Koh} is that 
 the monodromy representations of the braid group that 
arise on 
the solutions of the Knizhnik-Zamolodchikov (KZ) equation, 
which governs
the chiral WZNW conformal blocks, are equivalent to 
corresponding representations generated by the 
universal R-matrix of $U_q(\G)$. 
This aspect of the connection between quantum groups and 
WZNW models is 
most elegantly summarized by Drinfeld's construction  
of a canonical quasi-Hopf deformation $\A_h(\G)$ of 
the universal 
enveloping algebra $U(\G)$ that encodes the 
monodromy properties
of the KZ equation \cite{Dr1,Dr2}.
As a coalgebra $\A_h(\G) = U(\G)[[h]]$,  it has a 
quasi-triangular 
structure given by   $R_h = \exp\left(h \Omega\right)$, 
where $\Omega$ is 
the appropriately normalized tensor Casimir of $\G$, and 
a coassociator 
$\Phi\in \A_h(\G)\otimes \A_h(\G)\otimes \A_h(\G)$ 
defined by means of the KZ equation.
An advantage of the category of quasi-Hopf algebras is 
that it admits a notion of twisting, under which many 
different looking objects could be equivalent.  In particular,   
for generic $q$ $U_q(\G)$ is twist equivalent to $\A_h(\G)$,
which in many ways is a simpler object to consider.
For a review, see e.g.~\cite{CP}.

The classical limits of the Hopf-algebraic formal 
deformations of $U(\G)$ are the Lie bialgebra 
structures on $\G$, which integrate to Poisson-Lie groups,
while in the quasi-Hopf case the analogous objects are 
Lie quasi-bialgebras \cite{Dr2} and 
quasi-Poisson-Lie groups \cite{YKS}.
There have been many studies devoted to the description 
of the possible Poisson-Lie symmetries of the so called 
chiral WZNW phase space given by
\be
\M_{chir}:= \{\, g\in C^\infty(\R, G)\,\vert\, 
g(x+2\pi)=g(x)M, \quad M\in G\,\},
\label{Mchir}\ee
where $G$ is a connected Lie group corresponding to $\G$.    
The upshot (see e.g.~\cite{FG,BFP})
is that such a symmetry can be defined globally 
on $\M_{chir}$  whenever an appropriately
normalized solution $\hat r\in \G\wedge \G$  of 
the modified classical 
Yang-Baxter equation is available, which is used to define 
the Poisson-Lie structure on $G$ and appears
also in the definition of the appropriate
Poisson structure on $\M_{chir}$. 
In fact, the Poisson structure on $\M_{chir}$ can be encoded 
by a local formula of the following form: 
\be 
\Big\{g(x)\stackrel{\otimes}{,} g(y)\Big\}
={1\over\kappa }\Big(g(x)\otimes
g(y)\Big)\Big(
\hat r + \Omega \,{\mathrm sign}\,(y-x)
\Big), \quad 0< x,y<2\pi.
\label{xchPB}\ee 
Here $\Omega= {1\over 2} \sum_i T_i \otimes T^i$, where
$T_i$ and $T^i$ denote dual bases of $\G$ with respect 
to a symmetric, invariant, non-degenerate 
bilinear form $(\, \vert\,)$ on $\G$,
$(T_i\vert  T^j) =\delta_i^j$,
and  $\kappa$ 
is a constant.
In terms of the usual tensorial notation,
the Jacobi identity of (\ref{xchPB}) requires that
\be
[\hat r_{12}, \hat r_{23}] + 
[\hat r_{12}, \hat r_{13}]+ 
[\hat r_{13}, \hat r_{23}]=
- [\Omega_{12}, \Omega_{13}]\,. 
\label{mCYB}\ee
It is worth noting that for a compact simple Lie algebra
such an `exchange r-matrix' does not exist, because of
the negative sign on the right hand side of (\ref{mCYB}).

On the basis of the above mentioned 
results (see also \cite{Gab}), 
it is natural to expect that the classical analogue 
of the quasi-Hopf algebra $\A_h(\G)$ can 
also be made to 
act as a symmetry on $\M_{chir}$, but as far as we know this 
question has not 
been investigated so far.  Apparently, the  notion of a 
classical symmetry 
action of a quasi-Poisson-Lie group itself has 
only been defined
very recently \cite{AKS,AMKS}. 
The purpose of this letter is to point out 
that the expectation 
just alluded to is indeed correct, since the group $G$ 
equipped with the so called canonical quasi-Poisson-Lie
structure, which is the classical limit of $\A_h(\G)$, 
 acts as a symmetry in the sense of \cite{AKS} on the
chiral WZNW quasi-Poisson space.
We shall see that the simplest quasi-Poisson structure 
on $\M_{chir}$
for which this holds can still be 
defined by (\ref{xchPB}) but with $\hat r=0$.
This symmetry is thus available for any
Lie algebra (compact or not) with a invariant 
scalar product, since
it relies only on the Casimir $\Omega$.

In the next section we briefly recall  the necessary 
notions and then deal  with the quasi-Poisson 
structure of $\M_{chir}$ in sections 3 and 4. 
In the final section we  present
the natural generalization of certain results obtained
previously \cite{BFP} in the chiral WZNW example. 
 
\section{The notion of a  quasi-Poisson space} 

We first recall from \cite{Dr2,AKS} that
a Manin quasi-triple is a Lie algebra ${\cal D}$ with
a non-degenerate, invariant scalar product $\langle\,\,\vert
\,\,\rangle$ and
decomposition ${\cal D}={\cal G}\oplus{\cal H}$, where 
${\cal G}$ and ${\cal H}$ are maximal isotropic subspaces and
${\cal G}$ is also a Lie subalgebra in ${\cal D}$. $({\cal D},
{\cal G},{\cal H})$ becomes a Manin triple if ${\cal H}$
is a Lie subalgebra, too. Choosing a basis $\{e_i\}$ 
$i=1,2,\dots,n$ in ${\cal G}$ and corresponding dual 
basis $\{\varepsilon^j\}$ $j=1,2,\dots,n$ in ${\cal H}$
(satisfying $\langle e_i\vert\varepsilon^j\rangle=\delta_i^j$)
the commutation relations can be characterized in terms of the
structure constants $f_{ij\phantom{l}}^{\phantom{ij}l}$,
$F_{i\phantom{jl}}^{\phantom{i}jl}$ and $\varphi^{ijl}$
as follows:
\begin{eqnarray}
[e_i,e_j]&=&f_{ij\phantom{l}}^{\phantom{ij}l}e_l\,,
\label{comm1}\\
\left[e_i,\varepsilon^j\right]
&=&f_{li\phantom{j}}^{\phantom{li}j}
\varepsilon^l+F_{i\phantom{jl}}^{\phantom{i}jl}e_l\,,
\label{comm2}\\
\left[\varepsilon^i,\varepsilon^j\right]&=&
F_{l\phantom{ij}}^{\phantom{l}ij}\varepsilon^l+ 
\varphi^{ijl}e_l\,.\label{comm3}
\end{eqnarray}
Here $f_{ij\phantom{l}}^{\phantom{ij}l}$ are the structure
constants of the Lie algebra ${\cal G}$,
$F_{i\phantom{jl}}^{\phantom{i}jl}$ is antisymmetric in
$j,l$, $\varphi^{ijl}$ is totally antisymmetric;
and  the usual summation convention is in force. 
It is easy to write down the quadratic 
equations in terms of the
structure constants that express the Jacobi identities
of the commutation relations (\ref{comm1}-\ref{comm3}). One
of the equations (consisting of terms of the form $fF$) 
can be interpreted as requiring the
${\cal G}\rightarrow{\cal G}\wedge{\cal G}$ mapping $\hat F$
defined by $\hat F(e_i)=F_{i\phantom{jl}}^{\phantom{i}jl} 
e_j\otimes e_l$ to be a 1-cocycle. Further there are  
equations of $F\varphi$ and also of $FF+f\varphi$ type. 

The triple $({\cal G},\hat F,\hat\varphi)$ is called a 
Lie quasi-bialgebra,
where $\hat\varphi=\varphi^{ijl}e_i\otimes e_j\otimes e_l$ 
is here interpreted as a special 
element in $\wedge^3({\cal G})$.
It follows from (6) that
$(\G, \hat F, \hat\varphi)$ becomes a 
Lie bialgebra if $\hat\varphi=0$.
Note that $\hat F$ and $\hat\varphi$ correspond  
to the classical limits 
of the coproduct and the coassociator of a 
quasi-Hopf deformation of $U(\G)$.

A simple and important solution of the Jacobi constraints 
is given by $\hat F=0$ and
$\hat \varphi$ invariant with respect to the adjoint
action of ${\cal G}$. Below we shall use the
following \lq canonical' realization of this special case.
We consider a (real or complex) simple 
Lie algebra ${\cal G}$ and use
the invariant metric $\gamma_{ij}=(e_i\vert e_j)$ (where
$(\ \vert\  )$ denotes a fixed invariant scalar 
product on ${\cal G}$)
to raise and lower Lie algebra indices. ${\cal D}$ is given
by ${\cal G}\oplus{\cal G}$ and the scalar product is
defined by
\begin{equation}
\langle(a,b)\vert(c,d)\rangle=k\left[ (a\vert c)-
(b\vert d)\right]\,,
\label{k}\end{equation}
where $k$ is an arbitrary constant, which is real in 
the case of a real
Lie algebra. Now in terms of a
basis $\{T_i\}$ $i=1,2,\dots n$ in ${\cal G}$ the elements
of the dual ${\cal D}$ basis are written as
\begin{equation}
e_i=(T_i,T_i)\,,\qquad\qquad
\varepsilon^j=\frac{1}{2k}(T^j,-T^j)\,.
\label{evarepsilon}
\end{equation}
In this case $\hat\varphi$ is given by
\begin{equation}
\varphi^{ijl}=Cf^{ijl}\,,
\label{C}
\end{equation}
where $C=\frac{1}{4k^2}$. Note that while 
the Jacobi identities
for the commutation relations (\ref{comm1}-\ref{comm3})
are  algebraically satisfied with $\hat F=0$ and $\hat\varphi$
given by (\ref{C}) with any $C$, for a real Lie algebra 
${\cal G}$ the canonical realization defined above 
always gives $C>0$.

Manin quasi-triples and Lie quasi-bialgebras can be deformed
by the following transformation called \lq twist'. In terms
of the dual basis the twisted quasi-triple is defined by
$\tilde e_i=e_i$ and
\begin{equation}
\tilde\varepsilon^i=\varepsilon^i+t^{ij}e_j\,,
\label{twist}
\end{equation}
where $t^{ij}$ is an arbitrary  
antisymmetric matrix. The twisted structure constants
$f_{ij\phantom{l}}^{\phantom{ij}l}$,
$\tilde F_{i\phantom{jl}}^{\phantom{i}jl}$ and 
$\tilde \varphi^{ijl}$ are easily obtained by substituting
(\ref{twist}) into (\ref{comm2},\ref{comm3}). 
In particular, the 1-cocycle transforms as
\begin{equation}
\hat{\tilde F}(X)=\hat F(X)+[X\otimes 1+1\otimes X,\hat t\,]\,,
\qquad
\forall X\in \G,
\label{twistF}
\end{equation}
where $\hat t=t^{ij}e_i\otimes e_j$.

The importance of twisting is due to the fact that 
the canonical Lie quasi-bialgebra defined by the above example
can be twist transformed into a Lie 
bialgebra. In this case the 1-cocyle is a 1-coboundary,
\begin{equation}
\hat{\tilde F}(X)=[X\otimes 1+1\otimes X,\hat t\,]\,,
\label{twistFcan}
\end{equation}
and the requirement $\hat{\tilde\varphi}=0$ is equivalent
to the modified classical Yang-Baxter equation
\begin{equation}
[[\hat t,\hat t\,]]+C\hat f=0\,,
\label{mYB}
\end{equation}
where $[[\hat t,\hat t\,]]=[\hat t_{12},\hat t_{23}]+
[\hat t_{13},\hat t_{23}]+[\hat t_{12},\hat t_{13}]$
and $\hat f=f^{ijl}e_i\otimes e_j\otimes e_l$.
It is well-known that for a real compact Lie algebra
(\ref{mYB}) has solutions for $C<0$ only. Thus the
twisting from the canonical Lie quasi-bialgebra to
a Lie bialgebra is possible if ${\cal G}$ is a real
simple algebra for non-compact real forms only.
In the complex case it is always possible.

The global objects corresponding to Lie quasi-bialgebras 
$({\cal G},\hat F,\hat\varphi)$ 
are the {\em quasi-Poisson-Lie} groups \cite{YKS,AKS}. 
Such a Lie group with Lie algebra $\G$  
is equipped with a multiplicative bivector 
that corresponds to $\hat F$ 
and satisfies certain conditions involving $\hat{\varphi}$.
The precise definition is not needed in this letter.

The final notion to be recalled is that of a 
{\em quasi-Poisson space} \cite{AKS}.
This is a manifold ${\cal M}$ which is equipped with a
bivector $P_\M$ and on which a (left) action of a  
connected quasi-Poisson-Lie group $G$ is given subject 
to some conditions.
These conditions are specified below in terms of the 
corresponding Lie quasi-bialgebra.
To describe them,
note that on any manifold
a bivector can be encoded by the associated 
\lq quasi-Poisson bracket' $\{\,,\,\}$.
The quasi-Poisson bracket is associated 
with the bivector in the same way as
the Poisson bracket on a Poisson manifold is 
associated with the Poisson bivector \cite{MR}. 
Thus $\{\,,\,\}$ is a bilinear,
antisymmetric derivation on the functions on $\M$,
whose properties differ from the usual Poisson 
bracket only in that it does not necessarily satisfy the 
Jacobi identity. Using this language a quasi-Poisson space
is required to satisfy
\begin{equation}
\overline X_i\Big\{f,g\Big\}-
\Big\{\overline X_if,g\Big\}-
\Big\{f,\overline X_i g\Big\}=
-F_{i\phantom{jl}}^{\phantom{i}jl}
\Big(\overline X_jf\Big) 
\Big(\overline X_lg\Big) 
\label{Qequivar}
\end{equation}
and
\begin{equation}
\Big\{\Big\{f,g\Big\},h\Big\}+
\Big\{\Big\{h,f\Big\},g\Big\}+
\Big\{\Big\{g,h\Big\},f\Big\}=
-\varphi^{ijl}\Big(\overline X_if\Big)
\Big(\overline X_jg\Big)\Big(\overline X_lh\Big)\,.
\label{QJacobi}
\end{equation}
Here $\overline X_i$ are the infinitesimal generators of the
left action of the (quasi-Poisson-Lie) group $G$ on functions
on ${\cal M}$ satisfying $[ \overline X_i,\overline X_j]= 
f_{ij\phantom{l}}^{\phantom{ij}l}\overline X_l$.

It is easy to prove that if ${\cal M}$ equipped with
the quasi-Poisson bracket $\{\,,\,\}$ is a quasi-Poisson
space in the above sense with respect to 
some Lie quasi-bialgebra
then it is also a quasi-Poisson space with respect to the
twisted Lie quasi-bialgebra using the twisted bracket
\begin{equation}
\widetilde{\Big\{f,g\Big\}}=\Big\{f,g\Big\}
-t^{ij}\Big(\overline X_i f\Big)
 \Big(\overline X_jg\Big)\,.
\label{twistbarcket}
\end{equation}
Note that the requirements (\ref{Qequivar}) and (\ref{QJacobi})
ensure that the quasi-Poisson bracket can be restricted to 
the space of invariant functions where it becomes a genuine 
Poisson bracket, which is also invariant 
under twisting \cite{AKS}.

The above conditions become the usual conditions of a 
Poisson-Lie action on a Poisson manifold if $\hat\varphi=0$.
Another simplification occurs in the case of $\hat F=0$,
that is e.g. for the canonical structure, since in this case
(\ref{Qequivar}) simply means that the bivector $P_\M$ 
must be invariant under the action of $G$. 

\section{$\M_{chir}$ as a quasi-Poisson space}

After this short review we turn to the study of our 
example, the chiral WZNW phase space $\M_{chir}$.
We here show that this phase space can be equipped with
a very simple and natural quasi-Poisson bracket which
makes it a quasi-Poisson space with respect to the
canonical Lie quasi-bialgebra. In this example $G$
has to be identified with the WZNW group and the 
corresponding (left) action on functions of $g(x)$
is infinitesimally generated by
\begin{equation}
\overline X_ig(x)=g(x)T_i\,.
\label{INF}
\end{equation}
It is natural to equip $\M_{chir}$ with such a bivector which
guarantees that the components of the current 
$J_i(x)=\kappa (g'(x)g^{-1}(x)\vert T_i)$  generate 
the affine Kac-Moody algebra and
$g(y)$ is a primary field of this algebra:
\begin{equation}
\Big\{J_i(x),g(y)\Big\} = -T_ig(x)\delta (x-y)\,.
\end{equation}
The motivation for this second property is that it ensures 
the analogous property  of the full $G$-valued WZNW  field 
of which we are here considering the chiral part.
Now the simplest {\em Poisson bracket}  on $\M_{chir}$ 
consistent with the requirements that the
classical KM algebra relations together 
with the primary field nature
of  the chiral WZNW field  are 
reproduced is given by (\ref{xchPB}). Analogously, the
simplest {\em quasi-Poisson bracket} still consistent with
the above two requirements is
\begin{equation} 
\Big\{g(x)\stackrel{\otimes}{,} g(y)\Big\}
={1\over\kappa }\Big(g(x)\otimes
g(y)\Big)\,
\Omega \,{\mathrm sign}\,(y-x)
\,, \quad 0< x,y<2\pi,
\label{Qxch}
\end{equation} 
which is simply the exchange 
relation (\ref{xchPB}) with $\hat r=0$. 
Similarly to the Poisson brackets \cite{BFP}, 
the quasi-Poisson brackets are only defined 
for a class of \lq admissible' functions, which are
typically smeared out functions of the chiral WZNW field
$g(x)$ and the relation (\ref{Qxch}) has to be interpreted
in a distributional sense.

Now it is very easy to see that the equivariance
requirement (\ref{Qequivar}) (with $\hat F=0$) is satisfied
by the bracket (\ref{Qxch}) so we are left with (\ref{QJacobi}).
Since (\ref{Qxch}) is understood in a distributional sense,
we keep the arguments
$x,y$ and $z$ strictly different from each other and  then find
\begin{equation}
\Big\{\Big\{g(x)\stackrel{\otimes}{,}g(y)\Big\}
\stackrel{\otimes}{,}g(z)\Big\}\quad+\quad
\hbox{cycl.  perm.\,} 
=-\frac{1}{4\kappa^2}g(x)\otimes g(y)\otimes g(z)\,\hat f\,.
\label{QjacobiWZ}
\end{equation}
By taking into account (\ref{C}) and (\ref{INF}),
we conclude by comparing (\ref{QJacobi}) 
with (\ref{QjacobiWZ}) that 
$\M_{chir}$ with the bracket (\ref{Qxch}) 
is a quasi-Poisson space
with respect to the canonical Lie quasi-bialgebra if 
the constant $C$ in (\ref{C}) is 
chosen as $C=\frac{1}{4\kappa^2}$.
We make this choice by identifying the Lie quasi-bialgebra 
parameter $k$ in (\ref{k}) with the classical 
KM level parameter $\kappa$.

As we have mentioned, a canonical quasi-Poisson structure may 
be twisted to a genuine Poisson structure. In our example the
twist that transforms (\ref{Qxch}) into (\ref{xchPB})
is given by
\begin{equation}
\hat t=-\frac{1}{\kappa}\hat r\,.
\label{tw}
\end{equation}
Recalling (\ref{mYB}) and using $C=1/4\kappa^2$ we notice that
this transformation is possible if $\hat r$ satisfies the
modified classical YB equation $[[\hat r,\hat r]]=-(1/4)\hat f$.
Thus we see once more that $\M_{chir}$ can be equipped
with a genuine Poisson structure of the form of (\ref{xchPB}) for
complex or non-compact real 
groups only. On the other hand, the quasi-Poisson structure
(\ref{Qxch}) is perfectly well-defined for compact groups as well.

It is interesting to note that (\ref{Qxch}) is not the only
possible choice for a quasi-Poisson structure on $\M_{chir}$.
An other solution is defined as follows. The quasi-Poisson
brackets are still generated by the formula (\ref{xchPB})
but the constant exchange r-matrix $\hat r$ is replaced
by a \lq dynamical' 
r-matrix $\hat r_{\mathrm dyn}(M)$ depending on
the dynamical variables through the monodromy matrix $M$.
The $M$-dependence is given by the formula
\begin{equation}
r_{\mathrm dyn}(M)
=-\frac{1}{2}\tanh\left({1\over 2}{\mathrm ad} Y\right)\,,
\label{dynamicalr}
\end{equation}
where $Y:=\log M$ varies in a neighbourhood of zero in $\G$ on which 
the exponential parametrization of the monodromy matrix, $M=e^{Y}$, can be used.
In (\ref{dynamicalr}) $r_{\mathrm dyn}(M)$ has to be interpreted
as a linear operator on ${\cal G}$ corresponding to
$\hat r_{\mathrm dyn}(M)$ by means of the natural identification.

${\cal M}_{\rm chir}$ equipped with 
this \lq dynamical' quasi-Poisson structure is also a
quasi-Poisson space with respect to
the canonical Lie quasi-bialgebra, with $C=1/\kappa^2$,
at least on that subspace of the chiral WZNW phase space
where the exponential parametrization of $M$ is valid.
There it can also be twisted to a genuine Poisson
structure by adding a constant r-matrix, but now 
the constant r-matrix defining the twist (\ref{tw})
has to be normalized according to $[[\hat r,\hat r]]=-\hat f$.
The resulting twisted dynamical r-matrix, 
\begin{equation}
r+r_{\mathrm dyn}(M)
=r-\frac{1}{2}\tanh\left(\frac{1}{2}{\mathrm ad} Y\right) 
=r+\frac{1}{2}
\coth\left(\frac{1}{2}{\mathrm ad} Y \right)-
\coth\left({\mathrm ad} Y \right)\,,
\end{equation}
is the $\nu =1$ member of the family of dynamical r-matrices
constructed in (section 5 of)
\cite{BFP}, which were shown to satisfy a 
generalization of the modified classical YB equation and were 
interpreted in terms of certain Poisson-Lie groupoids.
We shall return to these dynamical r-matrices in section 5. 

\section{The monodromy matrix as momentum map}

In this section we would like to point out that the 
quasi-Poisson structure (\ref{Qxch}) on $\M_{chir}$ 
admits a group valued momentum map, which is provided by the
monodromy matrix of $g(x)\in \M_{chir}$, as might be expected.
We first recall the definition of the
generalized momentum map introduced in \cite{AKS}.
For this we need to consider the connected Lie groups $D$ 
and $G\subset D$ integrating 
${\cal D}$
and ${\cal G}$ respectively as well as the coset space $S=D/G$.
The action of $D$ on itself by left multiplications 
induces an action of $D$ 
on the coset space $S$. We denote the infinitesimal generators
of this left action on functions on $S$ by $X_i$ and $Y^j$
corresponding to $e_i$ and $\varepsilon^j$ respectively.
In the `canonical example' of our interest 
${\cal D}={\cal G}\oplus {\cal G}$,
and thus we can represent $D=G\times G$ as $D=\{(g_1,g_2)\}$
with $g_1,g_2\in G$ and the elements of the form $(g,g)$
give the subgroup $G$ embedded diagonally. Then the coset space
can be represented by $\{\sigma=g_1g_2^{-1}\}$, and hence $S$
in this case is identified with the group $G$ itself.
As a consequence of (\ref{evarepsilon}), 
 the infinitesimal generators  
of the $D$-action on $S\simeq G$ 
are explicitly given in the canonical case by
\begin{equation}
X_i\sigma=-[T_i,\sigma]\,,\qquad\qquad
Y^j\sigma=-\frac{1}{2k}[T^j,\sigma]_+\,.
\label{infinitesimal}
\end{equation}
Here $T_i$ and $\sigma$ are matrices in some linear 
representation corresponding to the
Lie algebra generators and the group element respectively
and $[\,,\,]_+$ means matrix anticommutator.

The momentum map
$\mu :{\cal M}\rightarrow S$ introduced in 
\cite{AKS} is required to satisfy the following two conditions.
First, it must be equivariant with respect to the 
corresponding $G$-actions,
which means that 
\begin{equation}
(X_i \Psi)\circ\mu=\overline X_i(\Psi\circ\mu)
\label{mom1}
\end{equation}
for any function $\Psi$ on $S$.
The second condition may be formulated as the equality 
\begin{equation}
\left\{\Psi\circ\mu ,f\right\}=
(\overline X_if)\, (Y^i \Psi)\circ\mu\,,
\label{mom2}
\end{equation}
where $\Psi$ is an arbitrary function on $S$ and $f$
an arbitrary function on ${\cal M}$. 
More precisely, one also requires a non-degeneracy 
condition, which ensures that (\ref{mom2}) can be solved 
for $\overline X_if$ in terms of the quasi-Poisson brackets of $f$ 
with the coordinate functions of $\mu$;
so that $\mu$ generates the vector fields
$\overline X_i$ on $\M$ through $\{\,,\,\}$. 
For the precise details the reader may consult \cite{AKS}.
There it is also shown that the momentum map $\mu$ is
always a bivector map, i.e. it satisfies
\begin{equation}
\left\{\Psi_1\circ\mu,\Psi_2\circ\mu\right\}=\left\{
\Psi_1,\Psi_2\right\}_S\circ\mu \,,
\label{mom3}
\end{equation}
where $\left\{\,,\,\right\}_S$ is the quasi-Poisson bracket
on the coset space $S$, which is a quasi-Poisson space itself
\cite{AKS}:
\begin{equation}
\left\{\Psi_1,\Psi_2\right\}_S=(Y^i\Psi_1)\,(X_i\Psi_2)\,.
\label{PBonS}
\end{equation}

In the canonical example, we can rewrite 
(\ref{mom1}) and (\ref{mom2}) more explicitly as 
\begin{equation}
-[T_i,\mu]=\overline X_i\mu
\label{mom1can}
\end{equation}
and
\begin{equation}
\left\{\mu,f\right\}=-\frac{1}{2k} (\overline X_i f) [T^i,\mu]_+\,,
\label{mom2can}
\end{equation}
where we use some matrix representation
of the elements of $S\simeq G$ like in (\ref{infinitesimal}).

Now our point is that for the quasi-Poisson
structure (\ref{Qxch}) on $\M_{chir}$
the momentum map is given by 
\be
\mu: \M_{chir} \ni g(x)\mapsto M=g^{-1}(x)g(x+2\pi )\in G\simeq S.
\label{Mmom}\ee
In fact, (\ref{INF}) implies immediately that 
$\mu=M$ satisfies (\ref{mom1can}).
Furthermore, one easily obtains from  (\ref{Qxch}) the relation
\begin{equation}
\left\{M\stackrel{\otimes}{,}g(x)\right\}=-
\frac{1}{2\kappa}[T^i,M]_+\otimes g(x)T_i\,,
\label{Mg}
\end{equation}
which is nothing but (\ref{mom2can}) in our case after 
the identification $k=\kappa$.
We conclude that the monodromy matrix plays the role of the group
valued momentum map with respect to the 
quasi-Poisson structure (\ref{Qxch}), 
similarly to 
its well-known \cite{FG} role in the 
Poisson-Lie context for the Poisson bracket 
given by (\ref{xchPB}).
Incidentally, the non-degeneracy condition mentioned above
is satisfied upon restriction to a 
domain of $\M_{chir}$, where $M$ is 
near enough to the unit element.

\section{Dynamical twists from quasi-Poisson to Poisson spaces}

Let $\hat {\cal R}\in \G\wedge \G= {\cal R}^{ij} T_i \otimes T_j$ 
be a constant classical r-matrix subject to
\be
[[ \hat {\cal R}, \hat {\cal R}]] = -\nu^2 \hat f,
\label{nuYB}\ee
where $\nu$ is an arbitrary constant.
Associate with $\hat {\cal R}$ a coboundary Lie bialgebra 
structure on $\G$ given by the cocommutator $\hat {\cal F}$ that 
operates as   
\be
\hat {\cal F}: T_i \mapsto 
{\cal F}_{i\phantom{jl}}^{\phantom{k}jl} T_j\otimes T_l
\quad\hbox{with}\quad
{\cal F}_{i\phantom{jl}}^{\phantom{i}jl}:={1\over k}\left(
{\cal R}^{jm} f_{mi}^{\phantom{mi} l} +
{\cal R}^{ml} f_{mi}^{\phantom{mi} j}\right),
\label{calF}\ee
where $k$ is an other arbitrary constant. 
A result of \cite{BFP} shows that locally it is possible to modify
the quasi-Poisson bracket (\ref{Qxch}) by 
shifting it with a monodromy dependent r-matrix in such a way 
to obtain a proper Poisson bracket
for which the natural left action of $\G$ 
becomes an {\em infinitesimal Poisson-Lie symmetry
with respect to the Lie bialgebra} $(\G, \hat {\cal F})$. 
We next recall these `dynamical twists' of the WZNW quasi-Poisson
space, and then point out their natural generalization.

For our construction we choose a 
neighbourhood of zero  $\check \G\subset \G$ 
which is diffeomorphic to an open submanifold 
$\check G \subset G$ by the exponential map.
We introduce a function 
$r_{\cal R}: \check \G\rightarrow  {\mathrm End}(\G)$  
with the aid of the formula
\be
r_{\cal R}: \check \G\ni Y \mapsto {\cal R} + 
 {1\over 2} 
\coth\left({1\over 2} {{\mathrm ad}Y}\right) - \nu
\coth \left(\nu {\mathrm ad}Y\right).
\label{rR}\ee
Here $(r_{\cal R}-{\cal R})$ is 
defined by means of the power series
expansion of the corresponding complex analytic 
function, $f_\nu(z)$,
around 
zero\footnote{For $\nu=0$ the complex function
$f_\nu(z):= {1\over 2} 
\coth\left({1\over 2} z\right) - \nu \coth (\nu z)$
becomes 
$f_0(z)= {1\over 2} \coth\left({1\over 2} z\right) - {1\over z}$.}, 
and $\check \G$ is chosen so that 
the corresponding power series in ${\mathrm ad}Y$ 
is convergent.
${\cal R}\in {\mathrm End}(\G)$ is the operator 
corresponding to $\hat {\cal R}$, and 
$\hat r_{\cal R}(Y)=r_{\cal R}^{ij}(Y) T_i \otimes T_j$
below denotes  the 
$\G\wedge \G$ valued dynamical r-matrix 
associated with $r_{\cal R}(Y)$.

By using the above dynamical r-matrix, 
we now define a Poisson bracket $\{\ ,\ \}_{\cal R}$  
on the domain of the chiral
WZNW phase space where the monodromy matrix $M$ 
varies in $\exp\left(\check \G\right)$.
This is given by  the 
following `dynamical twist' of (\ref{Qxch}):
\begin{equation} 
\Big\{g(x)\stackrel{\otimes}{,} g(y)\Big\}_{\cal R}
={1\over\kappa }\Big(g(x)\otimes
g(y)\Big)\Big(\,
\Omega \,{\mathrm sign}\,(y-x)
+ \hat r_{\cal R}(\log M)\,\Big)
\,, \quad 0< x,y<2\pi.
\label{shiftWZ}
\end{equation} 
We proved in \cite{BFP} that $\{\ ,\ \}_{\cal R}$ satisfies
the Jacobi identity and 
that (17) defines  an infinitesimal
Poisson-Lie symmetry with respect to 
$\{\ ,\ \}_{\cal R}$.
The second property means that the identity 
obtained from (14) by replacing $\{\ ,\ \}$ 
by $\{\ ,\ \}_{\cal R}$ and also 
replacing $F_{i\phantom{jl}}^{\phantom{i}jl}$ by
${\cal F}_{i\phantom{jl}}^{\phantom{i}jl}$ in (\ref{calF}),
with $k:=\kappa$,
can be verified.

Now we generalize the preceding construction to
the class of quasi-Poisson spaces that contains
the chiral WZNW phase space as
a representative example.
For this we consider a quasi-Poisson space 
$(\M,\{\ ,\ \})$ with respect to a canonical Lie quasi-bialgebra
such that the  $\G$-action admits a group valued
momentum map $\mu: \M\rightarrow G$.
More precisely, we also assume that 
$\check \M:=  
\mu^{-1}\left(\exp(\check \G)\right)$ is a 
non-empty open submanifold in $\M$, 
where $\check \G$ is introduced in (\ref{rR}), 
and restrict our attention to $\check \M$.
In general, we define  a new bracket 
$\{\ ,\ \}_{\cal R}$ on $\check \M$ by the following formula:
\be
\Big\{f, g\Big\}_{\cal R} := \Big\{ f, g\Big\} + 
{1\over k} r_{\cal R}^{ij}(\log \mu)
\Big(\overline X_i f\Big) \Big(\overline X_j g\Big),
\label{dyntwist}\ee
where $f$, $g$ are functions on $\check \M$ and
$\overline X_i$ are the infinitesimal  
generators of the $\G$-action.
It can be proven that $\{\ ,\ \}_{\cal R}$ {\em satisfies the Jacobi
identity and the equation} 
\begin{equation}
\overline X_i\Big\{f,g\Big\}_{\cal R}-
\Big\{\overline X_i f,g\Big\}_{\cal R}-
\Big\{f,\overline X_i g\Big\}_{\cal R}=
-{\cal F}_{i\phantom{jl}}^{\phantom{i}jl} 
\Big(\overline X_j f\Big) \Big(\overline X_l g\Big).
\label{PLshifted}
\end{equation}
This means that the $\overline X_i$ generate 
an infinitesimal Poisson-Lie 
action of the Lie bialgebra $(\G, \hat {\cal F})$
on the Poisson space $(\check \M, \{\ ,\ \}_{\cal R})$.

The proof is obtained straightforwardly 
by combining the properties of the momentum map $\mu$ 
with the properties of the r-matrix $r_{\cal R}(Y)$.
We only note here that the equivariance property
\be
[ 1 \otimes T + T \otimes 1, 
\hat r_{\cal R}(Y) -\hat {\cal R}]=
{d\over d \tau} \hat r_{\cal R} 
( e^{\tau T} Y e^{-\tau T})\vert_{\tau=0}
\qquad \forall T \in \G,
\label{equivar}\ee  
gives rise to (\ref{PLshifted}), while 
the Jacobi identity for $\{\ ,\ \}_{\cal R}$ hinges on 
a certain   
dynamical Yang-Baxter equation satisfied 
by $r_{\cal R}(Y)$ as shown in \cite{BFP}.
Notice that for $\nu={1\over 2}$ 
we have $\hat r_{\cal R}=\hat {\cal R}$ by (\ref{rR}),
and thus the `dynamical twist' of the quasi-Poisson bracket  
defined by (\ref{dyntwist})  simplifies to a special case of 
the `twist' that appears  in (\ref{twistbarcket}).
This explains our terminology. 
Similarly to the constant twists (\ref{twistbarcket}),
for $\G$-invariant functions  
$\{\ ,\ \}_{\cal R}$ gives the same as $\{\ ,\ \}$.

We wish to remark that the dynamical twist
is also available if ${\cal R}=0$ and $\nu=0$, in which case
the Poisson-Lie symmetry becomes a `classical $\G$-symmetry'.
This special case of our general construction 
has been considered previously in \cite{AMKS}.
If ${\cal R}=0$, then $r_{\cal R}(Y)$ 
becomes the Alekseev-Meinrenken
solution of the classical dynamical Yang-Baxter 
equation \cite{AM},
which was independently found in \cite{BFP}
in the context of the dynamical twist 
construction in the chiral WZNW model. 
In this case it is known  \cite{AMKS} (see also \cite{BFP})
that  with respect 
to $\{\ ,\ \}_{\cal R}$ the map 
$J:= -k\log \mu : \check \M \rightarrow \G \simeq \G^*$
is an equivariant  momentum map in the classical sense,
and the natural inverse of the dynamical 
twist construction is then available, too.
Namely, one can always translate from 
a Poisson space with a classical $\G$-symmetry generated
by an equivariant momentum map, $J$, to a quasi-Poisson space with
respect to a canonical Lie quasi-bialgebra: just
twist the Poisson bracket `backwards' 
as can be read off from (\ref{dyntwist}) 
with the identification $- k \log \mu=J$. 

Finally, it is worth stressing  that 
by combining the last mentioned construction with the 
dynamical twists introduced in (\ref{dyntwist}), 
locally (i.e.\ on $\check \M$)  one  
can always convert a classical
$\G$-symmetry into an infinitesimal Poisson-Lie symmetry 
associated with any constant, antisymmetric
solution of (\ref{nuYB}).

\medskip
For definiteness, 
so far we have taken $\G$ to be a simple Lie algebra 
since the WZNW model is usually 
studied in this setting thanks to 
the relationship with the affine Kac-Moody algebras.
However, as a classical field theory the model is 
equally well-defined 
for any finite dimensional Lie group $G$ whose Lie algebra  
carries an invariant scalar product.
It is clear that $\M_{chir}$ with (\ref{Qxch})
is a quasi-Poisson space in the same manner in 
all these examples as well.
Of course, the dynamical twist construction is also valid
in this general context.

\medskip
\bigskip

\noindent{\bf Acknowledgements.}
This investigation was supported in part by the Hungarian 
National Science Fund (OTKA) under T030099, T029802, T025120 
and by the
Ministry of Education under FKFP 0178/1999, FKFP 0596/1999.

\end{document}